# Is the Mg-related GaN "blue luminescence" deep-level an MgO surface state?

Or Haim Chaulker, Yury Turkulets, and Ilan Shalish*

*School of Electrical Engineering, Ben-Gurion University, Beer Sheva 8410501, Israel.*

GaN is taking over from silicon in power electronics, but its density of interface states has yet to be adequately controlled. The major step turning GaN into a technological semiconductor was its p-type doping. Mg is currently the only p-type dopant in technological use in GaN. Its incorporation into the lattice is difficult, requiring a thermal treatment that only partially activates the Mg. To achieve moderate p-type doping requires high doses of Mg that mostly remain inactive – a potential for defects. These defects have mostly been studied by photoluminescence that cannot differentiate bulk from surface states. Here, we used an absorption-based spectroscopy, originally invented by the inventors of the transistor – surface photovoltage spectroscopy. The results seem to transform the picture radically as far as our understanding of the famous Mg-associated blue luminescence. We observe an optical transition from the valence band into a deep trap around 0.49 eV above the valence band, along with what appears to be a complimentary transition from the same trap into the conduction band peaking at 2.84 eV (precisely coinciding with the "blue luminescence" energy). The similar shape of the spectra, their complimentary energies within the bandgap, and their opposite nature (hole vs. electron trap), appear to be more than a coincidence suggesting an Mg-related *surface state*. We suggest that small amounts of surface-segregated Mg oxidize during the post-growth Mg-activation heat treatment forming a surface state. HCl etch is observed to affect the photovoltage at the "blue luminescence"-related energy. A surface treatment is unlikely to affect the bulk. The only case that could support existence of these blue-emitting centers deeper than the surface is when they decorate extended defects – a special case of surface states. Finally, we show that reported luminescence from pure MgO produces the same blue luminescence even at the absence of GaN.

## I. INTRODUCTION

As Si technology is reaching unsurmountable limitations to its further development, GaN has been establishing its foothold as the next microelectronic technologic material.[1] The major technological leap that started the present era of GaN was its p-type doping by Mg.[2,3] At present, Mg is the only p-type GaN dopant in technological use. Over the three decades that have elapsed, it has been established, both experimentally and theoretically, that Mg incorporation into GaN induces two relatively shallow levels close to the valence band and an additional parasitic deep level.[4] The deep level is most commonly observed in photoluminescence as a wide emission peak more or less over the energy range between 2.6 and 3.0 eV, and has been dubbed the blue luminescence (BL).[5] While the shallow levels have been thoroughly studied both experimentally and theoretically and their explanation fairly established,[6,7] the origin of the deep level has remained unclear. Both experimental and theoretical studies of the BL deep level seem to corroborate a model suggested by Kaufmann *et al*. of a recombination that takes place *in the bulk* between a deep donor and the $Mg_{Ga}$ acceptor.[8,9,10] The clear linkage of the blue luminescence to the introduction of the Mg dopant has led all the previous studies to look for a transition involving a *bulk defect*. In this paper, we report new findings on this BL-related deep level that suggest a radically different scenario than the one presently accepted.

One crucial question that has not been answered experimentally is whether the observed optical transition takes place in the bulk or on the surface of the crystal. As a matter of fact, this question has never even been asked, as it was quite obvious that if the transition is associated with the presence of a dopant evenly distributed in the bulk, then the transition naturally takes place in the bulk. Unsurprisingly, the ensuing ab-initio work looked for a candidate bulk scenario to explain the transition and indeed found one. Apparently, the possibility of a surface state has never been imagined. Is there a good reason to consider a Mg dopant-related surface state? And if so, is it possible to answer this question experimentally?

Why should we consider a surface state? Mg is not readily incorporated into GaN. It requires a thermal treatment to help it occupy the right place in the lattice, and even then, only 1% of it is actually activated at best.[11] Mg has also been found to segregate at extended defects.[12,13,14] The crystal surface is by far the most extended defect there is. Should we not expect at least a minor, limited-extent, segregation at the growth surface of the crystal? Indeed, Siladie *et al* observed using atom-probe tomography an increase in Mg concentration towards the surface of GaN:Mg nanowires.[15,16] In fact, there seems to be a strong driving force to cause Mg to float to the growth surface. Mg oxide is a fairly stable oxide having an enthalpy of formation of -601.7 kJ/mol. The difference from the enthalpy of formation of Mg nitride (+288.7 KJ/mol) must drive the Mg atom to the surface to react with oxygen.[17] Oxygen-induced diffusion of the Mg dopant was indeed reported in GaN light-emitting diodes.[18] The availability of oxygen depends on the vacuum level during the growth. Indeed, molecular beam epitaxy (MBE), which is typically carried out under ultra-high vacuum, produces GaN:Mg that shows no BL emission, while methods such as MOCVD/MOVPE do produce it.[4,19] These latter methods are typically carried out under rough vacuum at the millitorr range, which somewhat reduces the oxidation probability after all but does not eliminate it altogether. One would then expect that not all the Mg that segregates to the surface during the crystal growth would end up as fully oxidized. Indeed, it has been reported that the BL emission intensity considerably increases in annealed GaN:Mg samples compared with their "as-grown" intensity.[8,20,21] In growth of heterostructures, Mg has been observed to out-diffuse from Mg-doped layers into subsequently grown layers, deleteriously affecting the performance of p-channel heterostructure field effect transistors and p-gate high electron mobility transistors.[22,23]

The question is then whether it is possible to differentiate experimentally between bulk and surface optical transitions? Photoluminescence is clearly not capable of the task and neither do deep-level transient spectroscopy (DLTS) as well as most of the other methods used for monitoring deep-levels, except perhaps for one method, *surface photovoltage spectroscopy*. This method was invented by the inventors of the transistor[24] but has seen limited use despite its excellent capabilities, possibly due to relative difficulties in its use and the required knowhow.[25] Even when using surface photovoltage, it is not always straightforward to tell surface from bulk transitions. However, it seems that in this special case, we were lucky.

## II. MATERIALS AND METHODS

The GaN samples used in this study were 5 um thick heavily-doped p-type GaN:Mg and were grown by hydride vapor phase epitaxy (HVPE) on sapphire. The carrier concentration was determined using Hall effect to be $\sim 1 \cdot 10^{18}\ cm^{-3}$.

Photoluminescence was excited in air at room temperature using a 1 mW He-Cd laser (Meles-Griot Ltd.) lasing at 325 nm and acquired using a Newport MS257 spectrometer and a Si CCD detector.

Surface photovoltage was measured using a Kelvin probe (Besocke Delta Phi Gmbh). The sample was illuminated using a halogen light source monochromatized by a Newport MS257 spectrometer and filtered by order-sorting filters. At each wavelength step, the sample was illuminated for 2 min, following which electrical measurements took place. The steps were equally spaced in energy (5 meV). At each point, the acquired value was obtained by averaging 100 consecutive measurements. The total time to acquire each spectrum was 12 hrs and it followed a relaxation period of about 14 days during which the sample was kept in the dark inside the Faraday cage vacuum chamber and its contact potential difference was continuously monitored to ensure that the sample has reached equilibrium before spectral acquisition commences. All the photovoltage spectra were obtained under a constant photon flux illumination. The flux was maintained constant using a software-controlled slit at the light source. It is important to control the flux during the acquisition rather than to normalize the spectra after the acquisition, because the latter method assumes that the optical response is linear with the light intensity, while in practice, this is rarely the case. The measurements were carried out inside a stainless-steel ultra-high vacuum cryostat chamber under vacuum of mid $10^{-9}$ Torr at liquid nitrogen temperature (77 °K).

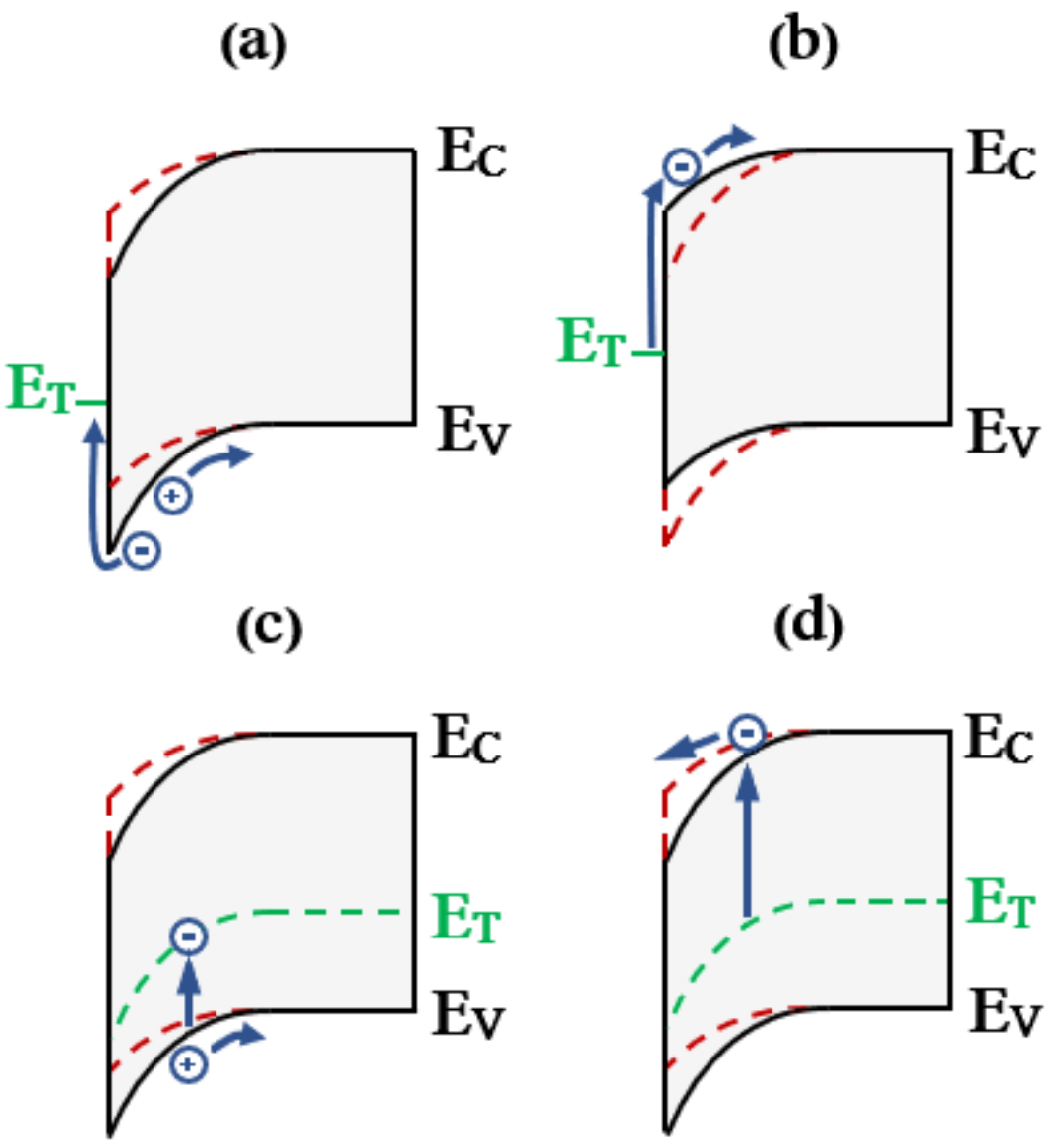

FIG 1 Cartoon showing four different optical transitions that generate surface photovoltage, i.e., in a change to the surface band bending. Cases (a) and (b) show transitions of an electron into, and from, a surface state, respectively. The resulting (changed) band bending is shown in red dashed lines. Cases (c) and (d) show transitions into, and from, a bulk state, respectively.

To understand our experimental approach, it is necessary to understand how the photovoltage responds in each of the following cases (we use a p-type material for this example without limiting the generality). When light of sub-bandgap wavelength is absorbed in a p-type semiconductor sample, there are several ways by which photovoltage may be generated.[26] Figure 1 shows 4 optical transitions that generate photovoltage. If the deep level is a surface state, the absorbed photon may either excite an electron from the valence band into the surface state (Fig. 1a), leaving a hole that is swept in the electric field into the bulk, in which case the surface band bending *decreases*, or excite an electron from the surface state into the conduction band (Fig. 1b) in which case the excited electron diffuses into the bulk against the electric field and the band bending *increases*. Hence, in the case of a surface state the polarity of the photovoltage signal clearly distinguishes between the two cases. If the deep level is in the bulk, the method can sense bulk transitions taking place within the surface space charge region. For a bulk state, two transitions are possible. A transition of an electron from the valence band into the state (Fig. 1c), or a transition from the state into the conduction band (Fig. 1d). In the former, the excited electron leaves behind a hole that is swept in the electric field into the bulk, while in the latter the excited electron will be swept in the conduction band by the electric field towards the surface. In both cases, the surface will be negatively charged resulting in a reduction of the surface band bending. Hence, in the case of a bulk state, the method *cannot* distinguish between the two types of transitions.

Here, we realized that this fundamental difference in the photovoltage response affords a means to distinguish between bulk and surface states. However, it is limited to cases where both types of transitions (from the valence band into the state, and from the same state into the conduction band), are possible and are actually observed.

To be able to observe both transitions (valence band to deep level and deep level to conduction band) for the same deep level requires that in equilibrium, part of the deep level distribution will be charged (i.e., occupied) and another part will be empty of charge (i.e., unoccupied). The transition from the valence band into the deep level should be into the unoccupied part of the distribution, which is typically the part that is away from the valence band, while the other transition (deep level to conduction band) should be from the occupied part which is typically closer to the valence band.

Finally, in the following figures, we show, beside the measured surface photovoltage, the *photon energy derivative* of the photovoltage. The derivative is proportional to the *density of the occupied, or unoccupied, states* within the deep level (depending on the type of transition) and appears as peaks somewhat similar to photoluminescence peaks, providing an easier means to interpret the results.[27] A major difference from photoluminescence is that photovoltage derivative spectra show also negative peaks. For surface states, a positive peak indicates an electron transition from an occupied state into the conduction band, while a negative peak indicates an electron transition from the valence band into an unoccupied state. Band-to-band (bandgap) transitions are positive for n-type conductive material and negative for p-type.

## III. RESULTS AND DISCUSSION

While the BL transition has been observed and studied so far by luminescence, it was studied here by surface photovoltage spectroscopy.[25] If we show photoluminescence in this study, we only use it as reference to show the existence of this luminescence or for comparison of peak position and no more. We did as best we could to *avoid* any interpretation based on photoluminescence as this was not intended to be a photoluminescence study – a spectroscopy based on absorption rather than emission.

### 3.1 Experiment I – Low temperature photovoltage

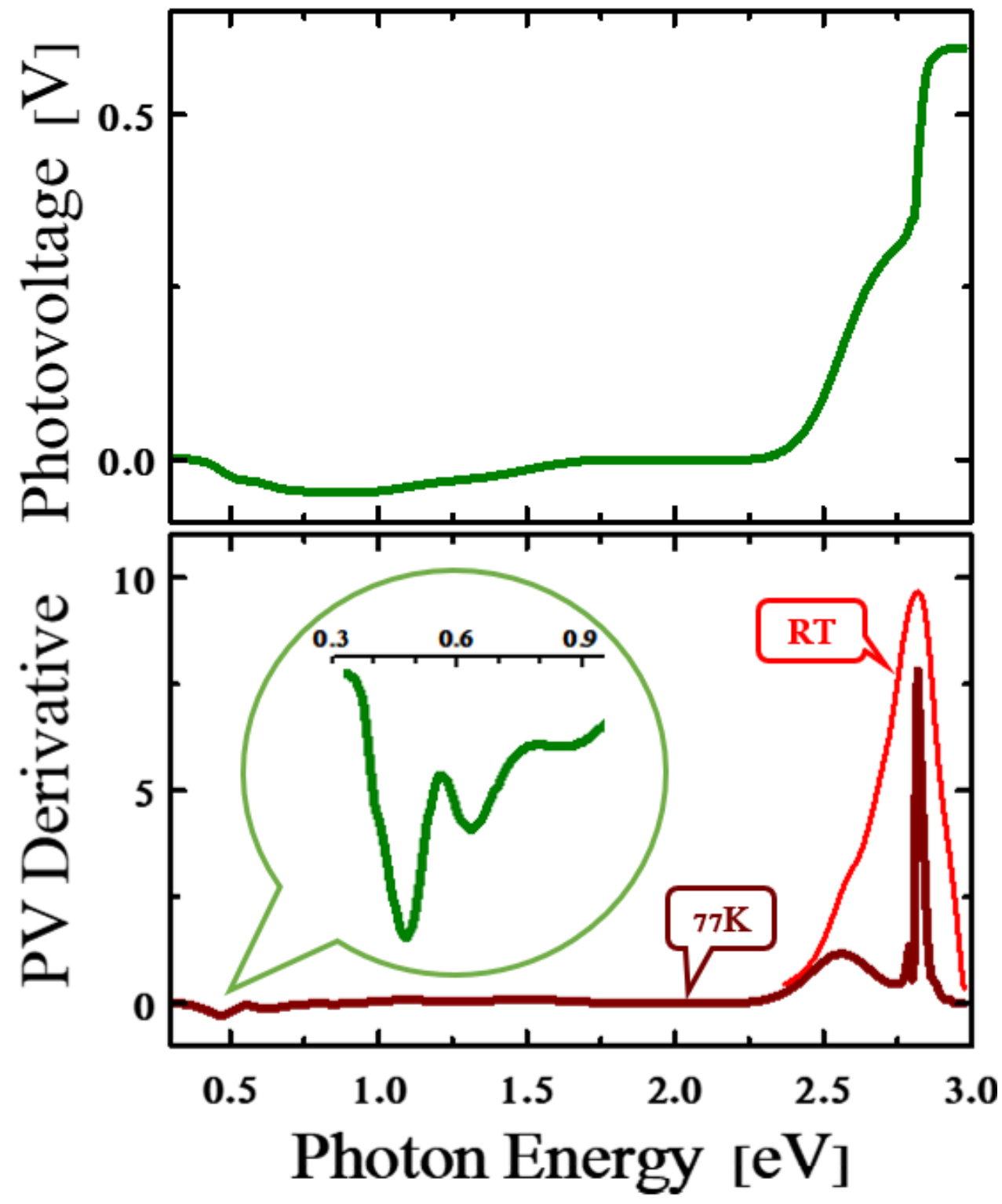

FIG 2 Surface photovoltage spectrum obtained from GaN:Mg sample (top) and its photon energy derivative (bottom). Changes of slope in the photovoltage spectrum indicate the onset of response of a deep level. They are more easily distinguished as peaks in the derivative spectrum that is also proportional to the density of occupied or unoccupied states in the deep level. In our case the negative peak at 2.84 eV indicates an electron transition from the *occupied* part of the deep level into the conduction band, while the positive peak at 0.49 eV (magnified in the inset) indicates an electron transition from the valence band into the *unoccupied* part of the deep level.

In the first experiment, we measured surface photovoltage at room temperature in ultra-high vacuum to see if we can detect by photovoltage the optical transition that is typically observed as a luminescence peak at 2.84 eV. Since photovoltage is caused by absorption, the occurrence of an absorption-related phenomenon at exactly the same position with the photoluminescence emission (related to recombination) seems to suggest that the two opposite transitions of absorption and recombination take place between the same states at the same energy. This excludes for example the possibility of a Stokes shift or a Franck-Condon shift, wherein absorption and recombination are expected to take place at very different energies. The Franck-Condon principle was originally proposed to explain observed differences between absorption and emission spectra and hence, in principle, is not justified in the absence of such difference. Nonetheless, it is most commonly used to explain photoluminescence, even in cases where absorption studies have not been carried out. To the best of our knowledge, absorption-related spectroscopies have not been used on the BL before.

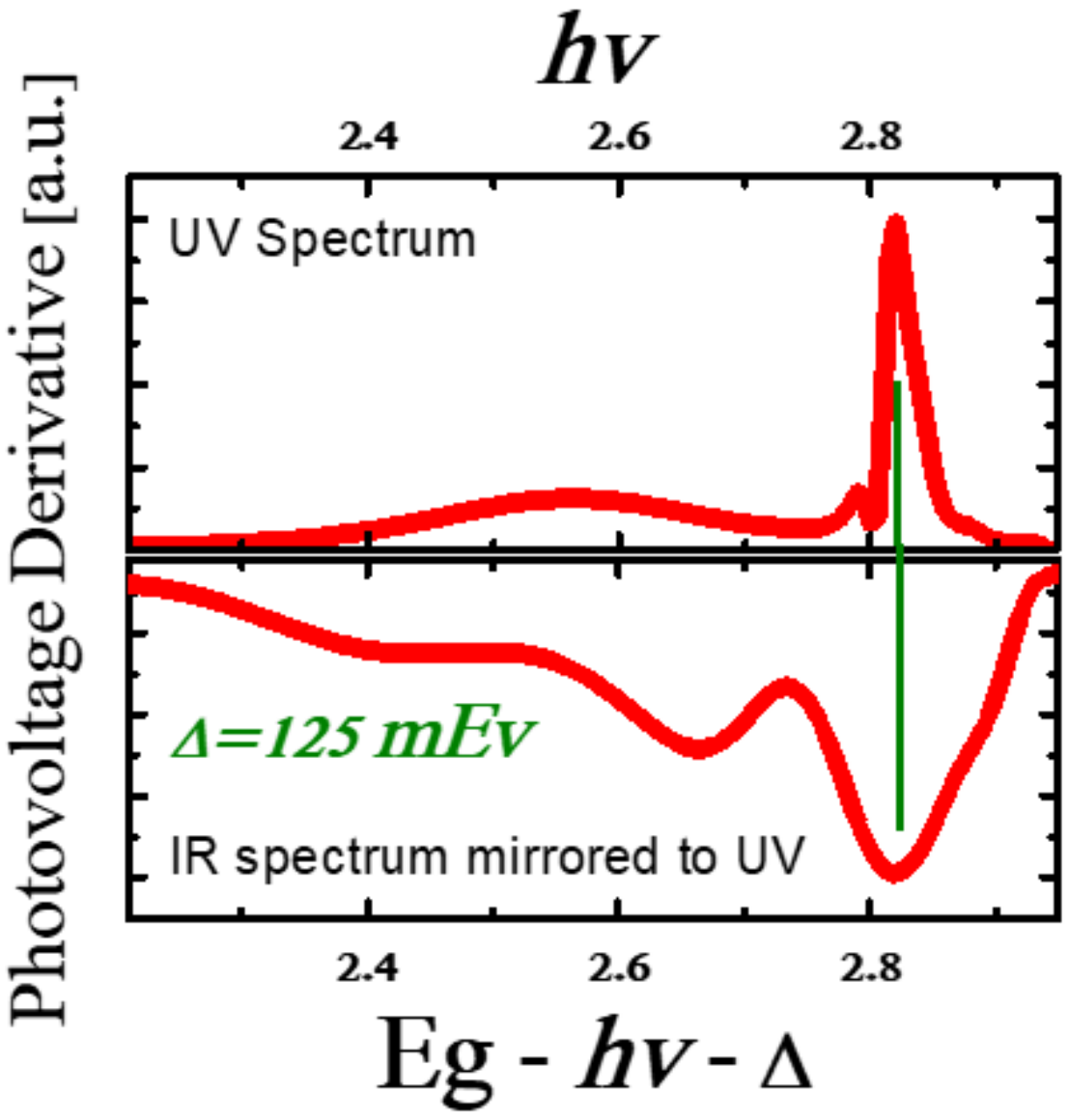

FIG 3 The photovoltage derivative in the bottom panel of Fig. 2 is replotted here. The top panel shows the UV range of the original plot as it is, while the bottom panel mirrors the IR part of the same spectrum into the same UV range by subtractive the photon energy values from the bandgap energy. Excluding a small difference of $\Delta = 125\ mEv$ the peaks appear to coincide. The small difference is probably because the transitions involve shallow donors and/or acceptors rather than the actual bands.

Of course, one may still go as far as to suggest that the absorption transition observed by photovoltage at 2.84 eV and the recombination transition observed by photoluminescence at 2.84 eV are two different and independent entities and only show up at exactly the same energy by mere coincidence, and that the recombination at this energy is still associated with absorption at a different energy according to the Franck-Condon principle. Such absorption would have to be observed. However, it has not. Therefore, it seems that *Occam's razor* would cut this complicated scenario in favor of the much simpler fact-based scenario proposed here. Moreover, to invoke the Franck-Condon principle, its *validity* to the case in hand has to be adequately verified.[28] To this end, the common practice is to compare the photoluminescence lifetime with the *lattice vibrational relaxation*.[29] However, the latter strictly relates to lattice vibrations *in the bulk* and does not apply to surface states at the very surface, where the periodicity of the lattice is discontinued.

In fact, surface states, being on the very surface of the crystal, are often electrically active and interact with molecules from air. [27,30] In order to be able to say anything about their vibrational time constant, a clear knowledge of their chemistry is required. As this is rarely possible, it seems that a valid argument can *rarely* be made on the validity of the Franck-Condon shift for surface states. To the best of our knowledge, this point has *never* been realized before.

Having observed an absorption-related phenomenon, photovoltage, at the exact BL peak position, we moved on to measure a wide range spectrum that included also the energy range, where one could expect a complementary transition, i.e., $Eg - 2.84 = 0.58\ eV$. As these peaks are typically wide, we set the range from 0.3 to 3 eV (4133 to 413 nm). To reduce thermal noise at the infrared, the sample was cooled to liquid nitrogen temperature, 77 °K.

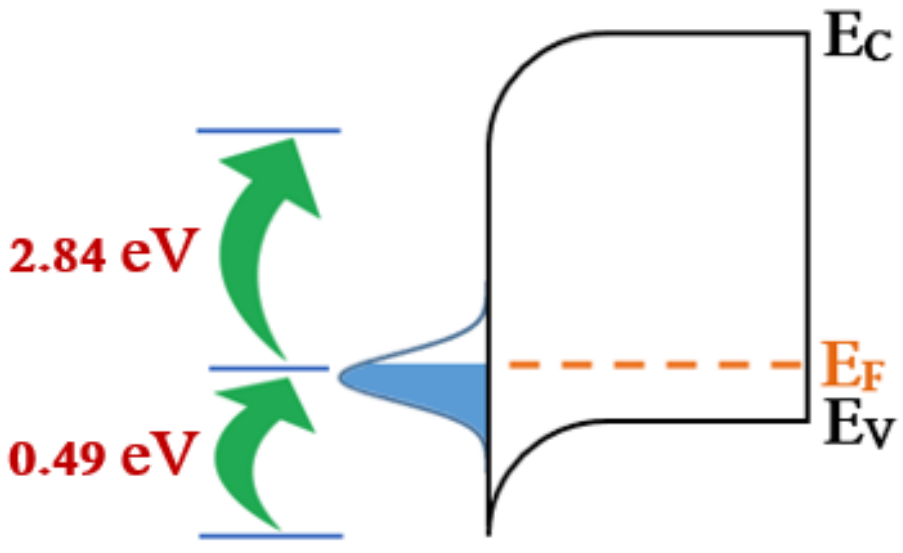

FIG 4 Cartoon showing our proposed explanation for the photovoltage observations. It depicts a p-type semiconductor having a surface state at its lower part of the bandgap. The state pins the Fermi level at the surface. The part below the Fermi level is occupied and the part above it is unoccupied. Electrons optically excited from the valence band into the unoccupied part result in a negative photovoltage (bands change their bending upward), while electrons excited from the occupied part into the conduction band cause a positive photovoltage (bands change their bending downward).

The top panel of **Fig. 2** shows the surface photovoltage spectrum thus obtained. The photovoltage is basically an integral over the transferred charge density.[27] The bottom panel shows the photon energy derivative of this photovoltage, which is relative to the density of either charged or discharged states. Each peak represents either charged or discharged states, but never both, depending on the type of transition. A positive peak is observed around 2.84 eV. An additional derivative spectrum taken at room temperature is shown as well in this panel for comparison. It shows how this peak becomes narrow at the low temperature. The same spectrum shows a much weaker negative peak at 0.49 eV. The positions of these two peaks *do not* add to the exact energy of the bandgap. The small difference of $\Delta = 125\ mEv$ is probably because the transitions are not to the actual bands but more likely to shallow donor or acceptor states close to the bands. To compare the two complementary transitions, we mirrored the low energy transition to the high energy part of the spectrum and placed the two transitions on the same energy scale. This is presented in Fig. 3. The complementary transition energies and the opposite polarities of the photovoltage derivative strongly suggest that the two transitions are *complimentary and involve the same state* as illustrated in the band diagram cartoon of Fig. 4. The peak at 0.49 eV appears wider than the one at 2.84 eV. This difference in peak width probably reflects to some extent the ratio between the phonon energy and the transition energy, because the same trend is observed for the two spectra taken at the blue energy range at two temperatures (the peaks appear to widen as the phonon energy becomes closer to the optical transition energy). However, it may also reflect the ratio between the width of the unoccupied and the occupied parts of this distribution. Hence, it appears that this state is, for the most part, empty of charge.

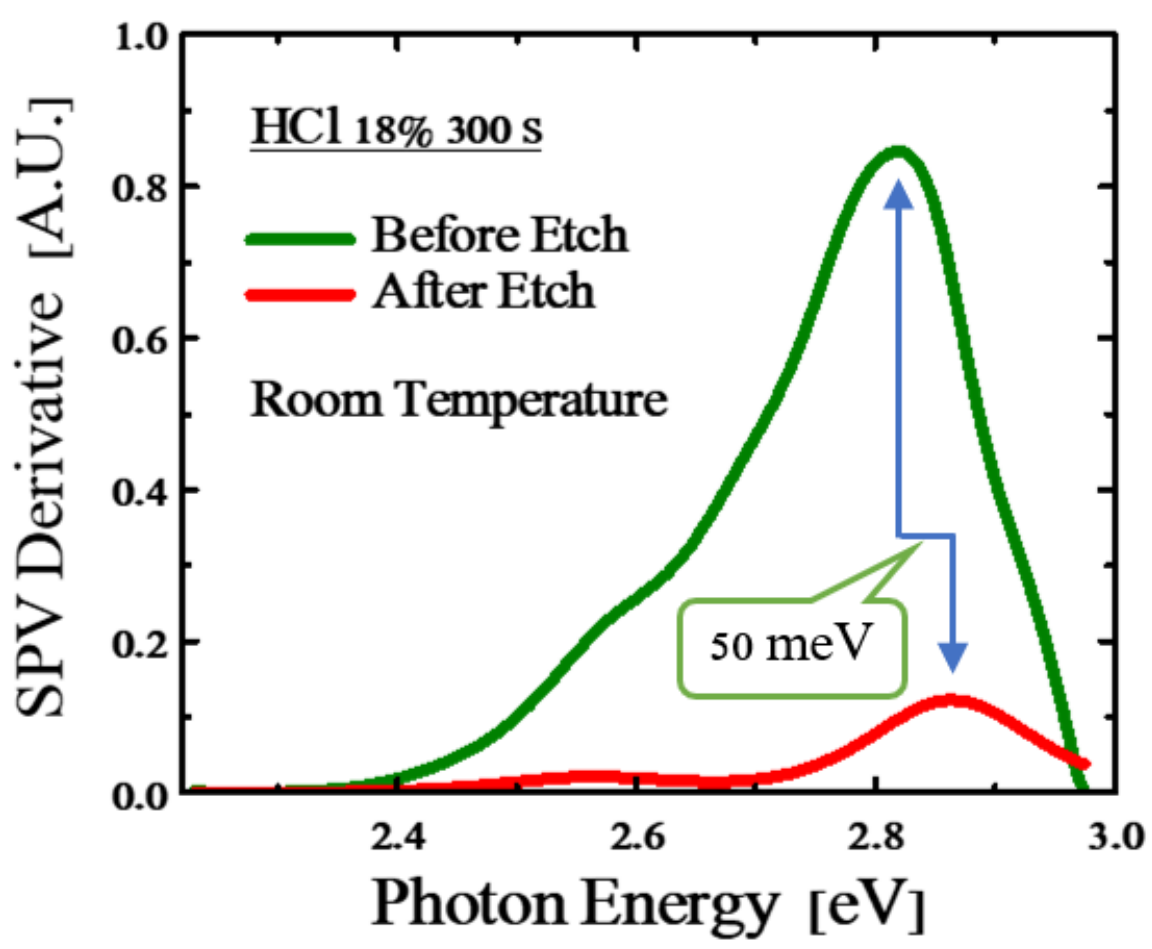

FIG 5 Photovoltage derivative spectra of a GaN:Mg sample before and after HCl etch. The spectra were acquired in vacuum at room temperature. The effect of the etch stands out clearly over the known range of the "blue luminescence" deep level, as expected. The peak area and height are clearly reduced by the etch. A blue shift of the peak by ~50 meV is observed after the etch as expected.

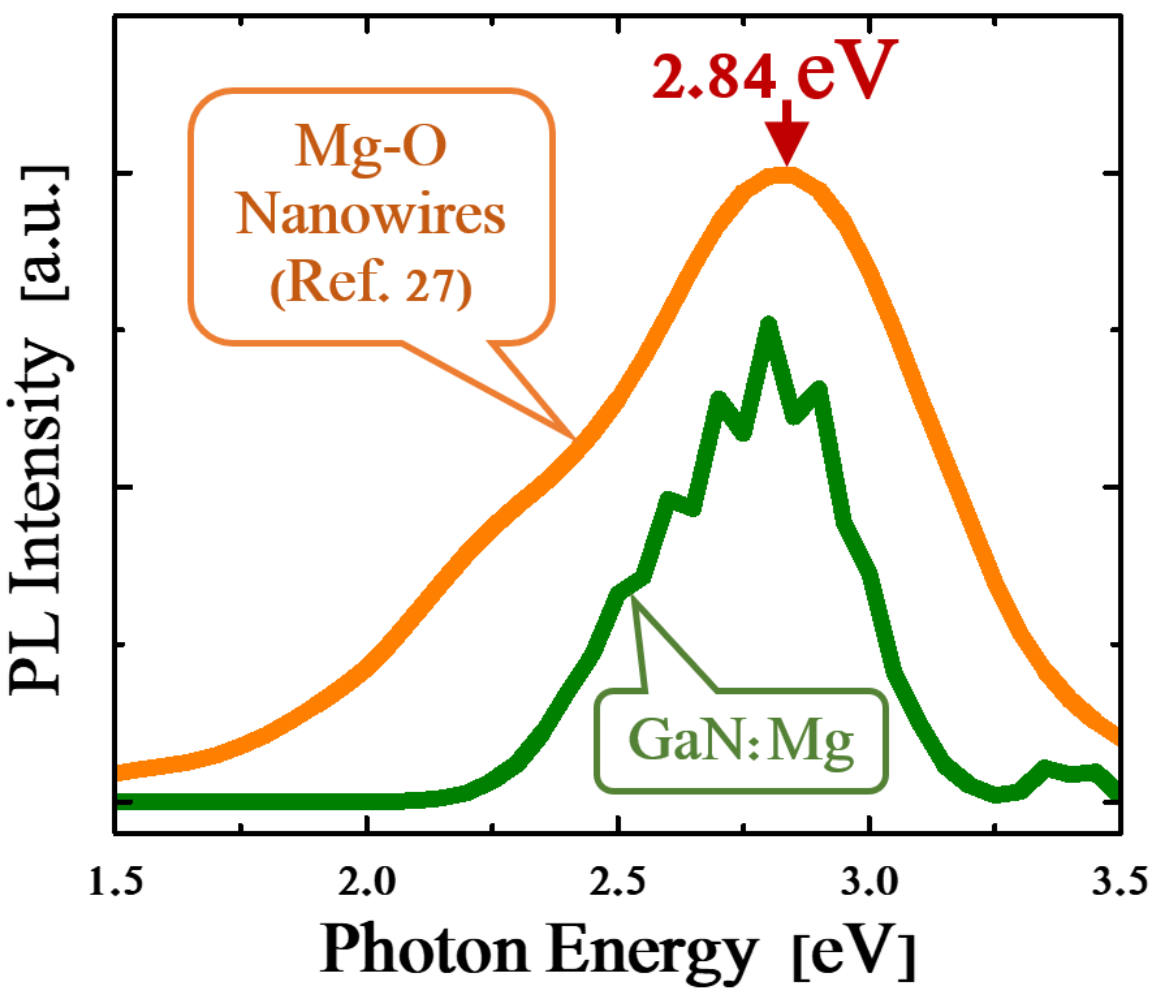

FIG 6 Photoluminescence spectra excited using a He-Cd laser at room temperature from a 2um-thick lowly-p-doped GaN:Mg grown by MOVPE on sapphire (green curve) compared with that of MgO nanowires (orange curve) reported in Ref. 27. Both spectra show a blue luminescence peak centered around 2.84 eV.

As previously discussed, this scenario is only possible if the involved state is a surface state. The photovoltage spectrum directly indicates that the 2.84 eV transition takes an electron from the charged part of the state and transfers it into the conduction band, or more precisely to a shallow donor close to the conduction band. This transfer positively charges the surface thus bending the bands down at the surface and hence causing a *positive* photovoltage. In contrast, the 0.49 eV transition takes an electron from the valence band (or more likely from a shallow acceptor) into the unoccupied part of this state distribution. This transition makes the surface more negatively charged, hence causing a *negative* photovoltage.

As explained in the experimental section (Fig.1), in order for two such complimentary transitions to give rise to different signs of photovoltage, the common deep level must be a surface state located about 0.5 eV above the GaN valence band. The polarity of the photovoltage transitions clearly and unambiguously indicate that this surface state is a deep acceptor, but this is not all. In order to have both an occupied (charged) part as well as an unoccupied (or empty) part, the Fermi level must divide this deep level distribution. However, in this heavily-doped p-type sample, the Fermi level in the bulk should be close to the valence band. Hence, this state can only be a *surface state*, where the bands are bent down and the Fermi level can be away from the valence band by as much as 0.5 eV. In other words, this surface state appears to pin the surface Fermi level.

A surface band bending of about 0.5 eV in a $1 \cdot 10^{18}\ cm^{-3}$ doped GaN should correspond roughly to a positive surface charge density of $\sim 2 \cdot 10^{12}\ cm^{-2}$. The total density of this surface state should be greater as it also includes the unoccupied part. Since only 1% of the Mg atoms become activated, to reach a carrier concentration of $1 \cdot 10^{18}\ cm^{-3}$, our wafer has to be doped with Mg concentration of $1 \cdot 10^{20}\ cm^{-3}$. If we consider the thickness of the surface to be the radius of the Mg atom, the density of dopants on the surface must be $1.6 \cdot 10^{12}\ cm^{-2}$.

This surface state interpretation raises a question: Can there be enough crystal volume associated with such surface state to support the typically strong luminescence observed? The answer is a definitive yes. It has been previously shown that surface state luminescence is associated with a surface layer as thick as the diffusion length of the generated electron-hole pairs. This would typically provide enough volume to support strong emission. A famous example of such surface emission is the commonly observed green luminescence in ZnO.[31] This transition was found to be affectively minimized by surface treatments and have been suggested based on several other methods to be associated with a surface state.[31,32,33,34,35,36] Nonetheless, its luminescence is typically rather strong.

### 3.2 Experiment II – Photovoltage before and after etch

If this is indeed a surface state, it should be vulnerable to surface treatments. Surface treatments cannot affect bulk states. If the observed transitions involve surface MgO, one should be able to etch at least some of the MgO using an acid. To this end, we tried 18% HCl for 300 s. For this experiment, we carried out a scan that concentrated on the "blue luminescence" range. As this range is in the visible part of the photon energy spectrum, low temperature was not required. Hence, the measurements were carried out at room temperature. As GaN is known to interact with the air, and in order to minimize this interaction to achieve equilibrium within a reasonable time, we carried out the measurements under ultra-high vacuum.

Figure 5 compares surface photovoltage derivative spectra before and after the etch. The intensity of the derivative is relative to density of charge. The effect of the etch stands out clearly over the known range of the "blue luminescence" deep level (2.4 to 3.0 eV), as expected. As surface Mg-O is etched away, the density of the surface states and their charge is greatly reduced and observed as a reduction in the peak height and area. The reduction in the surface charge causes a reduction in the built-in electric field within the surface space-charge region. This built-in field typically causes a certain red-shift in the peak position due to the Franz-Keldysh effect.[27,37] This shift increases with the strength of the built-in field. The reduction in surface charge reduces the field and causes the peak to shift back. Hence, the observed blue shift of 50 meV is to be expected, if the surface charge is indeed reduced.

If the BL were emitted from a deep level evenly spread in the bulk of the material, the charge in this state could not have been so majorly affected by etching. In a perfect crystal having no defects the etchant could only access the very surface. Today, GaN is still not perfect and still contains extended defects. While the density of these defects has reduced with the improvement of the growth methods, this problem is far from being totally eradicated. As mentioned before, Mg has been found to segregate to GaN extended defects.[12,13] It is also likely that oxygen diffuses inward from the surface more easily along extended defects to form Mg oxide. Hence, surface states emitting BL may form on extended defects at a considerable distance below the surface and thus be detected by methods such as depth resolved cathodoluminescence at a depth that may lead to an erroneous conclusion that these are bulk states. However, to be a true bulk state, a state has to be evenly spread in the bulk as is the case for, e.g., donor or acceptor doping impurities. Deep levels decorating an extended defect are not evenly spread in the bulk, and similarly to the formation of the surface depletion region due to charge trapped in surface states, the charge trapped in states decorating an extended defect causes a 3-dimentional depletion envelope surrounding the extended defect that behaves exactly as a surface state.[38] Hence to properly etch this surface state, the etchant should be able to somewhat diffuse inward at extended defects, preferably to the same depth that oxygen diffuses inward in these defects. Our HCl etch penetration appears to leave unetched some of the deep branches of those extended defect, as the density of the charge in this blue state does not diminish to zero.

### 3.3 Experiment III – Photoluminescence

If this surface state is indeed resulting from surface Mg-O, would pure Mg-O, without any GaN, show the same luminescence? To answer this question, we compare photoluminescence spectra from a lowly-doped GaN:Mg layer on sapphire with MgO photoluminescence from the literature.

Here, again, we stress that we avoid interpretation of photoluminescence experiments. We also refrain from judging, accepting, or rejecting, any of the interpretations or mechanisms proposed in those papers. We are only interested here in the plain observation of a photoluminescence peak at the relevant energy position.

Several papers report luminescence peaks exactly at the BL range from MgO single crystals. [39,40,41] **Figure 6** shows this comparison to photoluminescence of MgO nanowires reported in Ref. 41. Both spectra do show blue luminescence peaks with remarkable coincidence. This emission of MgO seems to have been overlooked by the GaN community.

While observation of such peak-position-coincidence alone may not be sufficient, it definitely supports a plausible scenario in which Mg, already known to segregate from GaN, segregates in rather small quantities to the GaN surface (likely through extended defects), where it inevitably oxidizes to produce a certain form of surface MgO, not necessarily a continuous layer, possibly limited to the vicinity of extended defects, wherein the BL transition takes place. This is in line with the observed effect of the etch in experiment 2 above. A mild HCl etch at room temperature does not attack the nitride, but would reduce a surface oxide and will do so only on the surface or in extended defects. This surface effect is again in line with the observation in experiment 1 above that may only be interpreted in terms of a photon-induced electron transition involving a surface state taking place at the BL transition energy.

## V. CONCLUSION

GaN is currently the next technological semiconductor after Si. If it ever makes it to become number one, it will only be after a method to reduce the density of surface states is found. To develop such a method requires knowledge of the GaN surface states. The present results shed new light on a well-studied deep level of p-type GaN, previously unknown to be a surface state. They seem to rest the case of the surface origin of the "blue luminescence" commonly observed in Mg-doped GaN.

While the transition is definitely related to the presence of Mg doping in the sample, it *appears not* to be induced by the Mg dopant but rather to be a transition taking place between MgO, likely formed in oxidation of surface (or extended defect)-segregated Mg, and the GaN host. While a study of the structure and composition of the surface was not attempted at this time, Mg segregation in GaN has been known and previously reported, and agrees well with the present results.

The conclusions that emerge seem to be compatible only with the premise that the GaN BL-related deep level is a surface state *likely* caused by MgO on the GaN surface.

While we show a distinct effect of a surface treatment on this state alone, the treatment used here was rather mild. Adequate elimination or passivation of this surface state calls for further trials and tuning to minimize the undesired byproduct of etch-induced surface damage.

**Acknowledgment**

Financial support from the Office of Naval Research Global through a NICOP Research Grant (No. N62909-18-1-2152) is gratefully acknowledged. Approved for public release DCN# 543-1495-24